\documentstyle[prl,aps,epsfig,multicol]{revtex}
\begin{document}
\topmargin 0.3cm
%\draft
\title{ Hysteresis and Spikes in the Quantum Hall Effect}
\author{ J. Zhu$^a$, H.L. Stormer$^{a,b}$, L.N. Pfeiffer$^b$, K.W. Baldwin$^b$, and K.W.
West$^b$}
\address{(a)Columbia University, New York, New York 10027\\
(b)Bell Laboratories, Lucent Technologies, Murray Hill, New Jersey
07974 }
\date{\today}
\maketitle

\begin{abstract}
We observe sharp peaks and strong hysteresis in the electronic
transport of a two-dimensional electron gas (2DEG) in the region
of the integral quantum Hall effect. The peaks decay on time
scales ranging from {\sl several minutes} to more than an {\sl
hour}. Momentary grounding of some of the contacts can vastly
modify the strength of the peaks. All these features disappear
under application of a negative bias voltage to the backside of
the specimen. We conclude, that a conduction channel parallel to
the high mobility 2DEG is the origin for the peaks and their
hysteretic behavior.
\end{abstract}
\begin{multicols}{2}

The hallmark of the integral and fractional quantum Hall effects
are wide regions of vanishing magnetoresistance and wide plateaus
in Hall resistance\cite{Girvin,Chakraborty,DasSarma}. These
features are centered around magnetic fields, that correspond to
integral or fractional fillings of Landau levels of a
two-dimensional electron gas (2DEG). Their origin is quantization
of the electron orbits into Landau levels and the formation of
localized states in real, slightly disordered 2DEG in the presence
of a high magnetic field, B. Electrons in the localized states
provide a reservoir, which is in equilibrium with the
current-carrying, delocalized states and keep their occupation
constant over wide stretches of B. While in the integral quantum
Hall effect (IQHE)\cite{Girvin} the ingredients of this picture
are of single-particle origin, they are of many-particle origin in
the fractional quantum Hall effect
(FQHE)\cite{Chakraborty,DasSarma}.

Two recent experiments have observed hysteretic behavior and/or
peak formation in electronic transport of 2DEG in the regime of
the FQHE\cite{Kronmueller,Cho}. Minor discrepancies in data taken
on opposite field sweeps are common and are usually attributed to
the large inductance of the magnet and the resulting time delays
or to slight temperature differences between both sweeps. However,
the recently observed effects are very large. Kronmueller \emph{et
al.}\cite{Kronmueller} observed the appearance of a huge spike at
the position of the $\nu=2/3$ minimum when the magnetic field is
ramped very slowly. The time scale for development of this feature
can be as long as hours, which suggested the involvement of
nuclear spins in its creation. Cho \emph{et al.}\cite{Cho} have
observed hysteretic behavior in resistance traces taken on several
fractions around filling factor $\nu=1/2$ and attribute it to
non-equilibrium phases of composite fermions in this regime. The
origin of these observations remains puzzling and the nature of
the underlying non-equilibria remains unclear.

We have observed strong hysteresis and the formation of sharp
peaks in magneto-transport experiments on 2DEGs in quantum wells
in the regime of the IQHE at temperatures of $\sim$0.1K. To our
recollection, we have never observed such features in a
traditional single heterojunction interface sample. The
characteristic decay times for the sharp peaks can be as long as
several hours. Their life time can be strongly altered by
momentary grounding of the contacts. Hysteresis and spikes
disappear on application of a voltage bias to an electrode on the
back side of the specimen. We conclude that the origin of
hysteretic behavior and spike formation in our samples is the
result of a non-equilibrium charge distribution, which arises due
to the coexistence and dynamic exchange of electrons between the
high-mobility 2DEG and a low-mobility parallel conduction channel
in the vicinity of the doping layer.

Our samples are modulation-doped GaAs/AlGaAs {\sl quantum well}
structures grown by molecular-beam epitaxy (MBE). A high density
2DEG resides in a 300\AA ~wide quantum well 2000\AA ~below the
surface. The well is $\delta$-doped on both sides with silicon
impurities at a distance of 950\AA ~in sample A and 750\AA ~in
sample B. Samples are cleaved into 4mm$\times$4mm squares and
eight indium contacts are diffused at the corners and the middle
of the edges. Transport measurements are performed using standard
lock-in techniques in a dilution refrigerator with a base
temperature of 70mK. A 100nA current is used in most of the
measurements. Both samples have mobility higher than
$13\times10^{6}cm^2$/Vsec. The density is
$2.3\times10^{11}cm^{-2}$ in sample A and
$3.2\times10^{11}cm^{-2}$ in sample B.

\begin{figure}
\begin{center}
 \epsfig{file=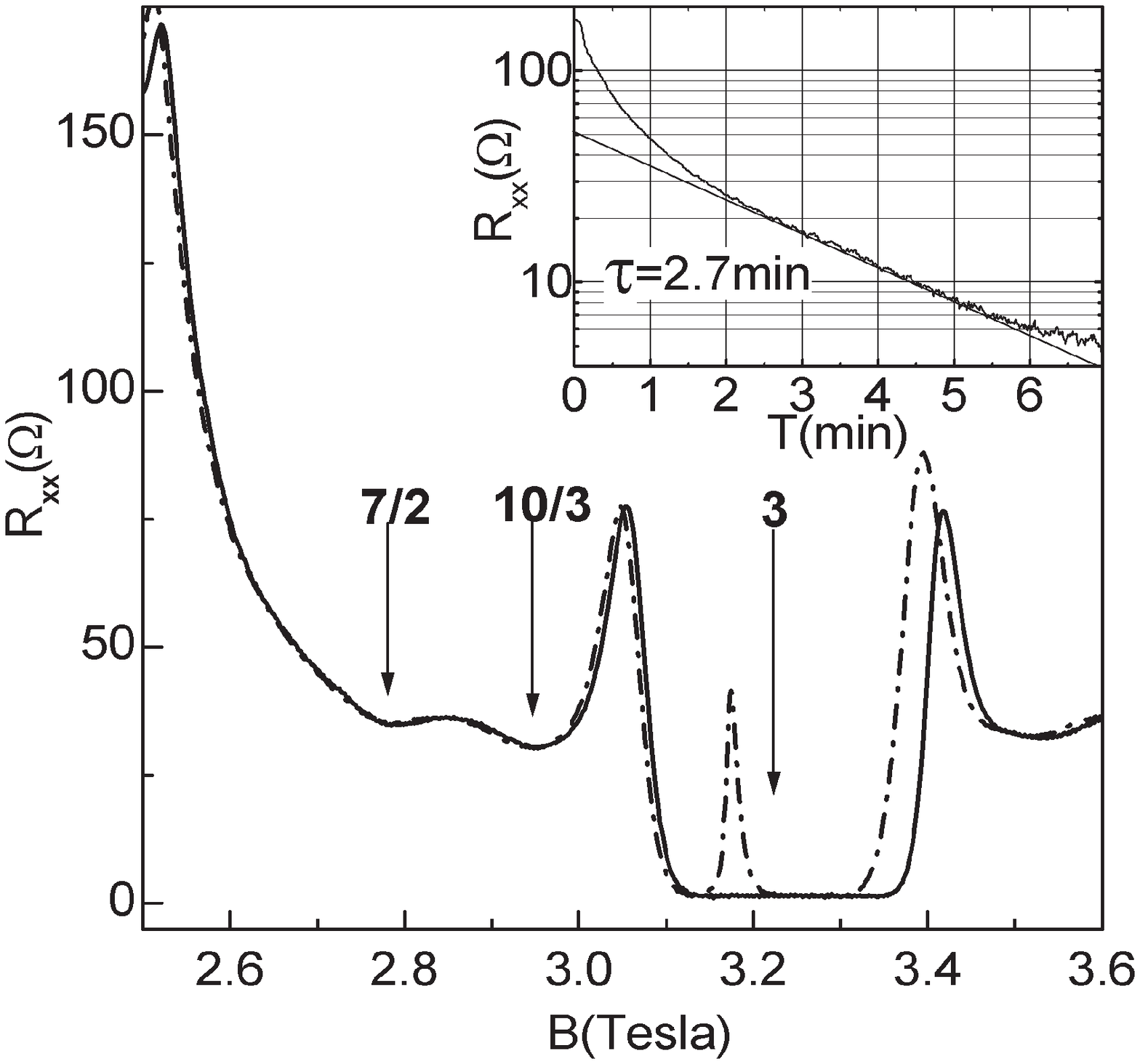, width=7in}
\end{center}
%\vspace{-2cm}
 \caption{Magnetoresistance of sample A in a magnetic field from 2.5T to 3.6T. Solid
 and dash-dotted traces represent upward and downward field sweeps, respectively. Arrows indicate
 filling factor $\nu$. A sharp hysteretic resistance peak occurs at $\nu=3$. Inset: A typical
 decay process of the peak at $\nu=5$.}
  \label{spikeat3}
 \end{figure}

As an example of the hysteresis and resistance spikes that arise
at many IQHE positions, we show in Fig.\ref{spikeat3} the
magnetoresistance, $R_{xx}$, of sample A in the vicinity of
filling factor $\nu=3$, measured at a slow sweep rate of
0.05T/min. Solid and dash-dotted lines represent traces taken on
upward and downward field sweeps, respectively. Both traces
largely reproduce, although there is a slightly discrepancy in the
position of the high-field flank. Most remarkably, however, a
sharp spike appears in the central part of the $\nu=3$ minimum.
This peak is only $\sim$0.01T wide, comparable in height to the
surrounding $R_{xx}$, and it is completely missing on the
up-sweep.

\begin{figure}
   \begin{center}
\epsfig{file=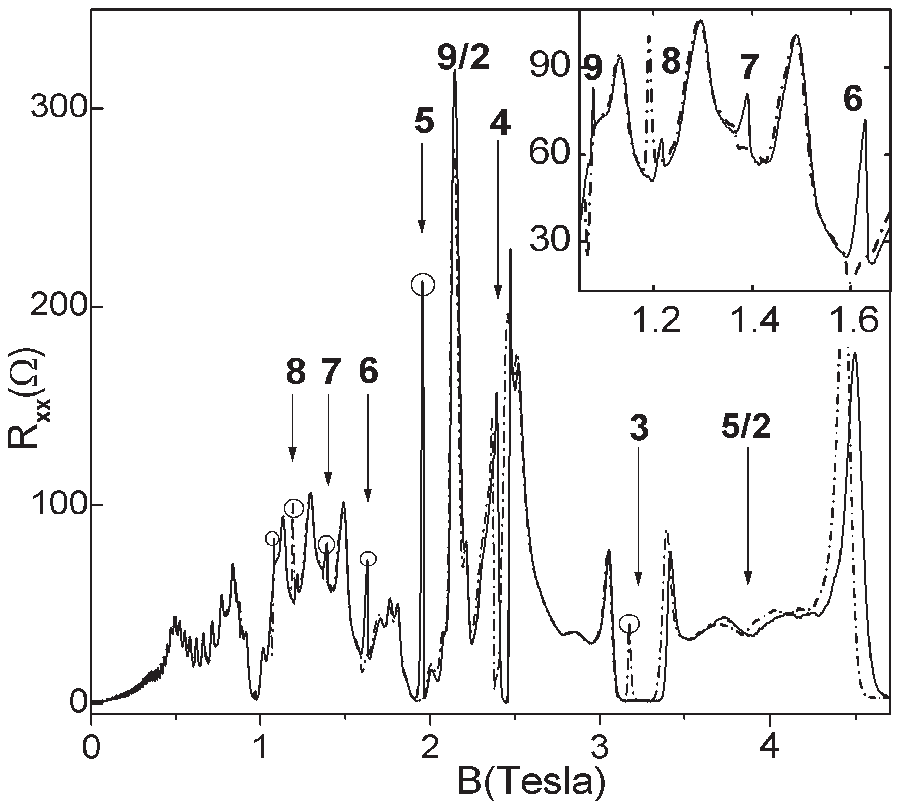, width=7in}
\end{center}
\vspace{-5cm}
 \caption{Magnetoresistance of sample A up to 4.7T. Notation
 is the same as in Fig.\ref{spikeat3}. Sharp resistance peaks are
 circled. Inset: Expanded scale of region between 1T and 1.7T,
 corresponding to $\nu$=6 to 9.}
 \label{morespikes}
 \end{figure}

Hysteretic resistance peaks similar to the one observed around
$\nu=3$ occur in our sample at the positions of most resolved
integral filling factors. Fig.\ref{morespikes} shows an extended
$R_{xx}$ traces of sample A. A superposition of two oscillations
is evident. A set of sharper Shubnikov-de Haas (SdH) oscillations
is superimposed on a slowly oscillating background. In spite of
this complication, we can clearly identify the positions of the
IQHE minima in the SdH oscillations, of which we have labeled
$\nu=3$ through 9 (see also inset on an expanded scale). Similar
to the occurrence of the sharp peak in the $\nu=3$ minimum, such
spikes are also present at these higher filling factors (see
circles), most notably at $\nu=5$, where the spike dominates over
any other features of this trace. The details of the hysteresis
vary. The peaks appear on the up-sweep ($\nu=9,7,6,5$) or on the
down-sweep ($\nu=3$) and in some cases in both directions, but
with very different amplitudes ($\nu=8$). Instead of being absent
in one direction, the previous peak position is sometimes marked
by a small dip. At fields, higher than $\nu=3$ and up to the
highest field of 14T (not shown), traces from both field sweeps
largely reproduce and there are no further spikes observed. In
this high-field regime all the regular features of the IQHE and
FQHE are well developed.

Sample B behaves similar to sample A, despite the difference in
electron density. The shapes of the low field envelopes in both
samples resemble one another. Their maxima and minima occur
roughly at the same field values. The backgrounds in both samples
reach their highest values near 2.5T and gradually disappear above
it. In sample A, the region of most pronounced hysteresis occurs
for $9\geq\nu\geq3$ while in sample B, this occurs for $12\geq
\nu\geq4$.

To examine the time dependence of the peaks, we swept the field
slowly to their maxima and stopped at the summit. The inset of
Fig.\ref{spikeat3} shows the subsequent time evolution of the peak
at $\nu=5$. After a rapid drop the decay becomes exponential with
a time constant of $\tau\sim$2.7min. The time constant is
sample-dependent, $\nu$-dependent and depends on the contact
configuration. The typical value for sample A is a few minutes and
for sample B, a few hours, with a maximum of $\sim$10 hours. These
are enormously long time scales for the decay of an electrical
resistance in a 2DEG.

We made a critical observation during such decay experiments in
sample B. A quick grounding and subsequent un-grounding of some of
the contacts led to a dramatic decrease of the amplitude of the
peaks. Although this observation was not reproducible during all
cool-downs, it points clearly to the existence of some
non-uniform, non-equilibrium configuration, that can be
equilibrated by the redistribution of charge within the specimen.

To characterize peak creation and decay we performed several
additional experiments. The amplitudes of the peaks increase with
increasing sweep rate, but much less than proportional. They
increase by less than a factor of two when the sweep rate is
raised twenty times from 0.02T/min to 0.4T/min. The decay of the
peaks with time probably accounts for the small difference in
amplitude. Raising the temperature weakens the observed hysteresis
and increases the background, while the amplitudes of the peaks
and dips shrink. The hysteresis at $\nu=5$ in sample A disappears
at about 400mK. Sample B shows similar behavior. Both AC and DC
current excitations were used to conduct the experiments and data
from both largely resemble each other. The data remain essentially
independent of AC current amplitude from 10nA to 1$\mu$A.

The oscillatory background in the low field data strongly suggests
the existence of a conduction path in parallel to the 2DEG
channel. The occupation of a second subband in the quantum well
can safely be ruled out on the basis of a simple calculation.
Another source for a parallel current path are the Silicon
modulation-doping layers on both sides of the quantum well. A
fraction of the electrons, can remain at the site of this layer
and provide a conducting path in parallel to the 2DEG. At high
enough mobility, such an impurity channel can exhibit its own
magneto-transport, superimposed on $R_{xx}$ from the 2DEG.

\begin{figure}
 \begin{center}
 %\vspace{-1.5cm}
 %\hspace{-1cm}
  \epsfig{file=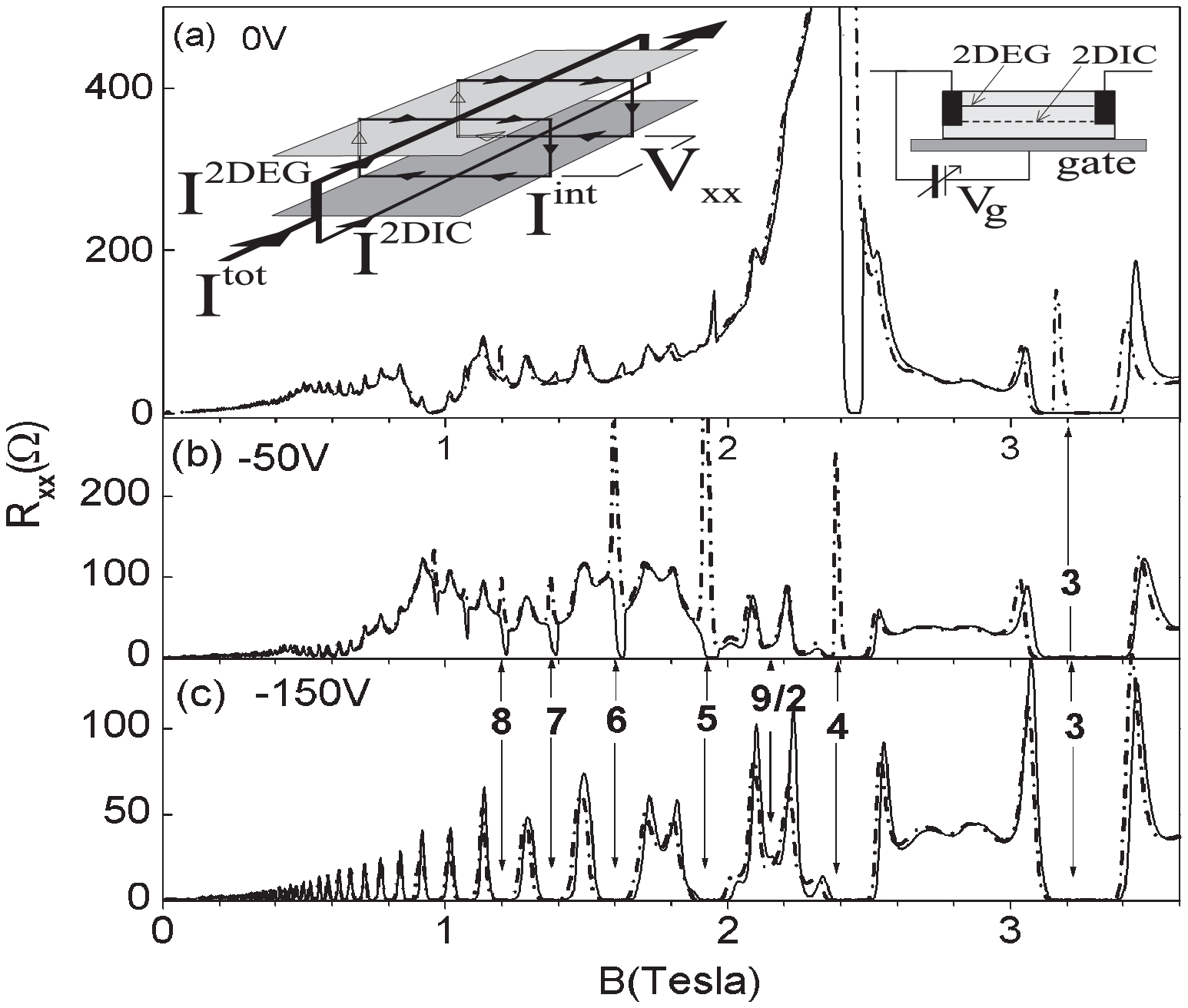,width=7.5in}
\end{center}
\vspace{-5cm}
 \caption{Magnetoresistance of sample A at different backgate
 voltages. Solid and dash-dotted lines represent upward and
 downward field sweeps, respectively. Left inset schematically shows the current configuration in the
two-channel system. Right inset shows the diagram of the sample
and the backside gate.}
 \label{backgate}
 \end{figure}

To investigate the origin of the parallel channel, we applied a
voltage, $V_g$, to a backside electrode (gate) placed under the
substrate. Fig.\ref{backgate} shows the magnetoresistance of
sample A at different backgate biases. Here we have chosen a
current direction perpendicular to the one used in
Figs.\ref{spikeat3} and \ref{morespikes}. Although the hysteresis
is less pronounced in this current configuration, several
hysteretic peaks are clearly visible in panel (a), at zero bias. A
negative voltage of -50V on the backgate across the 0.5mm thick
substrate has a dramatic effect on the trace in panel (b). While
the $R_{xx}$ background weakens, the previously small hysteresis
peaks grow enormously in amplitude and the sharp spikes at
$\nu=4,5$ and 6 dominate the graph. Uniformly, peaks occur on the
down-sweep, while the up-sweeps either show the customary IQHE
minima ($\nu=4,5$) or sharp downward cusps that approach vanishing
$R_{xx}$ ($\nu=6,7,8$). The last and cleanest hysteretic peak has
moved from its previous position at $\nu=3$ in panel (a) to
$\nu=4$ in panel (b). In general, with increasing back bias, we
see such a progression from higher to lower magnetic field.
Eventually, all hysteresis effects disappear in panel (c) at a
backgate bias of -150V. The specimen shows a clean $R_{xx}$
behavior as is customary for very high mobility samples. Not only
is the fragile states at $\nu=7/2$ visible, but the data also show
the recently discovered anisotropic state at $\nu=9/2$ and
11/2\cite{Lilly,Du}, manifested by deep minima in
Fig.\ref{backgate} (c) and well-documented clear maxima in
Fig.\ref{morespikes}.

Backgate bias does not change the electron density in the 2DEG as
is evident from the stationary B-positions of its IQHE and FQHE
features. The oscillations of the background, on the other hand,
are steadily moving to lower field. Sample B behaves similarly to
sample A and its data resembles those of panel (c) at a higher
bias of -170V. The backgate bias experiments on both samples
provide strong evidence for the existence of a parallel conducting
path in the form of a two-dimensional impurity channel (2DIC) on
the substrate-side of the quantum well and for its role in the
appearance of the background as well as the hysteretic spikes. The
symmetric doping channel on the top side of the sample does not
provide such a parallel channel, probably due to depletion by the
nearby surface of the sample.

From the shift of the minimum in the background at $\sim$0.95T in
Fig.\ref{backgate} (a) with increasing $V_g$ and a gate
capacitance of $C_g\sim22pF/cm^2$ we derive an initial density of
$n_{2DIC}\sim5.7\times10^{10}cm^{-2}$ in the impurity channel.
This identifies the 0.95T-minimum in Fig.\ref{morespikes} as the
$\nu=2$ IQHE of the 2DIC . The other minima in the background
follow quite well the usual 1/B sequence, with the strong minimum
at B$\sim$1.9T representing $\nu=1$. This indicates at least a
moderate mobility for this 2DIC, since spin-splitting is just
resolved. At -150V the density of the 2DIC has been depleted to
$3.0\times10^{10}cm^{-2}$ and has probably fallen below the
conduction limit of such a disordered channel. Therefore, parallel
conduction has vanished and a clean $R_{xx}$ trace is observed.

In the remainder of the paper, we develop a model that can account
for many of our observations. We regard our system as consisting
of two parallel sheets of conductor, a high-mobility 2DEG and a
low-mobility 2DIC. Both are connected via eight contacts at the
perimeter of the specimen and are coupled by a mutual capacitance
of $C=120nF/cm^2$. At any given magnetic field a complex current
distribution emerges. The situation with the 2DEG in a quantum
Hall state is shown as an inset to Fig.\ref{backgate}. We do not
differentiate between resitance, R, and resistivity, $\rho$, since
both differ only by a factor of order of unity in our square
sample. Following the value of the Hall resistances the total
current $I^{tot}$ divides between both layers according to their
density ratio, $n^{2DIC}/n^{2DEG}$. Therefore, about 1/5 of the
total current is flowing through the 2DIC. In addition, due to the
different resistivities between the 2DEG and the 2DIC, an {\sl
interlayer} current $I^{int}$ is induced, which contributes to the
voltage $V_{xx}$. Whenever the 2DEG is in the IQHE, a simple
calculation\cite{calculation} shows, that the measured
$R_{xx}=V_{xx}/I^{tot}$ simply reflects $\rho_{xx}^{2DIC}$,
attenuated by the square of the ratio of electron densities in
both layers. From a value of $\sim100\Omega$ for the smoothly
varying background in Fig.\ref{morespikes}, we deduce
$\rho_{xx}^{2DIC}\sim$2.5k$\Omega$. This establishes the parallel
channel as a moderately good conductor.

As the magnetic field is swept, the Fermi levels of both layers
oscillate due to Landau quantization, creating an oscillating
imbalance in the chemical potential between the layers. The
resulting potential difference is particularly drastic in the
regime of the IQHE of the 2DEG, where its Fermi energy changes
abruptly by $\hbar\omega_c$. To keep the Fermi levels in
equilibrium, a charge $Q\sim C\hbar\omega_c/e$ needs to be
transferred between the layers. At B$\sim3$T,
$Q\sim4\times10^9e/cm^2$ or $\sim10\%$ of the charge of the 2DIC.
The relevant series conductance to charge or discharge the 2DEG
sheet from its edge, where the contacts reside, is its diagonal
conductivity $\sigma_{xx}$. In the regime of the IQHE this
parameter tends toward zero. The resulting RC time constant for
equilibration can assume values as long as hours if
$\sigma_{xx}\sim10^{-11}\Omega^{-1}$, which is not uncommon in a
high mobility 2DEG sample\cite{Eisenstein,icps}.

The combination of a finite field sweep rate and a long time
constant gives rise to a density inhomogeneity in the 2DEG. While
the edges of the 2DEG and the 2DIC are quickly equilibrated, the
center of the 2DEG lags far behind and maintains a higher or lower
electron density concentration depending on the field sweep
direction. The resulting radial density gradient in the 2DEG is
imprinted with opposite sign onto the 2DIC due to the
electrostatic interaction between both layers. Since the diagonal
resistivity, $\rho_{xx}^{2DIC}$, in the low density, disordered
2DIC depends strongly on electron density, the sudden change of
the Fermi level in the 2DEG and the resulting density gradient in
the 2DIC can abruptly alter the local $\rho_{xx}^{2DIC}$ and hence
the current pattern. This leads to a non-equilibrium $V_{xx}$
which is observed in the experiment as a time-dependent spike. In
particular, if the electron system in the 2DIC is near one of the
metal-insulator transitions, as they arise close to the edge of a
Landau level, the change in $\rho_{xx}^{2DIC}$ and hence in
$V_{xx}$ can be dramatic. This explains why exceedingly sharp
spikes always occur on a very low or completely absent background
such as $\nu=3,5$ in Fig.\ref{morespikes}. It also accounts for
the enormous growth of the spikes with decreasing carrier density
in the 2DIC in Fig.\ref{backgate} (b). To predict the direction a
particular spike is pointing (up or down) requires knowledge of
the transient non-equilibrium current distribution. This pattern
can be very complex since it depends on the local
$\rho_{xx}^{2DIC}$. which is sensitive to the position of the
Fermi level with respect to the density of states in the
disordered 2DIC. The resulting current pattern is difficult to
assess in detail.

The characteristic time of the phenomenon is the RC time of the
electric discharge between 2DEG and 2DIC. However, it is {\sl not}
the actual discharge current that is observed in $V_{xx}$. There
is simply insufficient charge in the capacitor to account for the
observed, minute-long interlayer current. What is rather observed
is the influence of the induced non-equilibrium charge
distribution in the 2DEG on the resistivity pattern in the 2DIC
and the resulting transient redistribution of currents in the
specimen. The narrowness of the spikes is a result of the
narrowness of the regions within the IQHE over which $\sigma_{xx}$
takes on sufficiently low values to create sufficiently long RC
times to be observable on the time scale of our experiment.
Outside of these narrow regions of exceedingly small $\sigma_{xx}$
charge transfer between 2DEG and 2DIC happens rapidly, both layers
maintain equilibrium and $V_{xx}$ is time independent. Raising
temperature increases $\sigma_{xx}$, therefore, peaks disappear at
high temperature. The long decay times of the peaks is a direct
result of the long RC times. Our model also explains the
hysteresis of the spikes. Obviously, opposite field sweeps
generate radial density gradients of opposite signs which lead to
different current patterns and hence hysteretic behavior.

 In conclusion, the strong spikes and large hysteresis in the
magnetoresistance of our 2DEG in the regime of the IQHE are the
result of a non-equilibrium charge distribution caused by the long
RC times to modify the electron density in the 2DEG in the IQHE
regime. The origin of these spikes is a parallel impurity channel.
We can explain our observations as resulting from the capacitive
coupling between 2DEG and this neighboring impurity channel and
the time-dependent current distribution it creates.

% \vspace{-0.6cm}

\end{multicols}
%%%%%%%%%%%%%%%%%%%%%%%%%%%%%%%%%%%%%%%%%%%%%%%%%%%%%%%%%%%%%%%%%%%%%%
\newpage

  %\vspace{-0.3cm}

\begin{references}
\bibitem{calculation}
The total injected current $I^{tot}$ divides between both layers
according to their two-point resistances, which, apart from small
contact resistances, are $\rho_{xy}^{2DEG}$ for the 2DEG and $\sim
\rho_{xy}^{2DIC}+ \rho_{xx}^{2DIC}$ for the 2DIC. Assuming
$(\rho_{xy}^{2DIC})^2\gg(\rho_{xx}^{2DIC})^2$, we obtain
$I^{2DIC}/I^{2DEG}\sim n^{2DIC}/n^{2DEG}$. While the longitudinal
voltage drop $V_{xx}^{2DEG}$ in the 2DEG is zero by virtue of its
IQHE state, $V_{xx}^{2DIC}\sim \rho_{xx}^{2DIC}\times I^{2DIC}$.
This potential difference induces an interlayer current $I^{int}$
through the voltage contacts which increases $V_{xx}^{2DEG}$ by
$\rho_{xy}^{2DEG}\times I^{int}$ and reduces $V_{xx}^{2DIC}$ by
$\sim\rho_{xy}^{2DIC}\times I^{int}$\cite{Syphers}. In equilibrium
$I^{int}\sim I^{2DIC}\times \rho_{xx}^{2DIC}/\rho_{xy}^{2DIC}$
since $\rho_{xy}^{2DIC}\gg \rho_{xy}^{2DEG}$. This yields
$R_{xx}=V_{xx}/I^{tot}\sim\rho_{xx}^{2DIC}\times(\rho_{xy}^{2DEG}/\rho_{xy}^{2DIC})^2
\sim\rho_{xx}^{2DIC}\times(n^{2DIC}/n^{2DEG})^2\sim\rho_{xx}^{2DIC}\times(1/5)^2$.
\end{references}
\end{document}